\documentclass[letterpaper]{jpconf}
\usepackage{graphicx,epsfig}

\def\gtrsim{\lower.5ex\hbox{$\; \buildrel > \over\sim \;$}} 
\def\lesssim{\lower.5ex\hbox{$\; \buildrel < \over\sim \;$}} 

\begin{document}

\title{High Energy Cosmic Rays from Local GRBs}

\author{Armen Atoyan$^1$ and Charles D.\ Dermer$^2$}

\address{$^1\,$CRM, Universit\'e de Montr\'eal, Montr\'eal, Canada H3C 3J7;
\small atoyan@crm.umontreal.ca}
\address{$^2\,$NRL, Code 7653, 
Washington, DC 20375-5352, USA; \small charles.dermer@nrl.navy.mil}

\begin{abstract}
We have developed a model that explains cosmic 
rays with energies $E$ between $\sim 0.1-1\,\rm PeV$ and the energy of the 
second knee at  $E_2 \sim 3\times 10^{17} \,\rm eV$ as originating from  
a recent Galactic gamma-ray burst (GRB) that occurred  $\sim 1\,\rm Myr$ 
ago within $ 1\,\rm kpc$ from Earth. 
Relativistic shocks from GRBs are assumed to inject power-law distributions
of cosmic ray (CR) protons and ions to the highest ($\gtrsim 10^{20}\,\rm eV$)
energies. Diffusive propagation of CRs from a recent ($\sim 1\,\rm Myr$ old) 
GRB explains the CR 
spectrum near and above the first knee at $E_1 \sim 3\times 10^{15}\,\rm eV$. 
The first and second knees are explained as being directly 
connected with the injection of plasma turbulence in the interstellar 
medium on $\sim 1\,\rm pc$ and $\sim 100\,\rm pc$ scales, respectively.
 Transition to CRs from 
extragalactic GRBs occurs at $E\gtrsim E_2$. The origin of the ankle in 
the CR spectrum at $E_{\rm ank}\simeq 4\times 10^{18}\,\rm eV$ is due to 
photopair energy losses of UHECRs on cosmological timescales, as also 
suggested by Berezinsky and collaborators. The rate density of 
extragalactic GRBs is assumed to 
be proportional to the cosmological starburst activity 
in the universe. Any significant excess flux of extremely high energy CRs 
deviating from the exponential cutoff behavior at  
$E> E_{GZK}\simeq 6\times 10^{19}\,\rm eV$  would imply
a significant contribution due to recent GRB activity on timescales 
$t\lesssim 10^8\,\rm yrs$ from local extragalactic sources  
within $\sim 10\, \rm Mpc$.
\end{abstract}

\section{Introduction}

\vspace{2mm}

There is general consensus that acceleration of CRs by supernova remnants 
(SNRs) is the main contributor of galactic CRs at energies below 
$\sim 100$ TeV (e.g. \cite{drury,gp85}). It is also generally 
thought that all CRs with energies up to at least the second knee
in the CR spectrum at $E_2\sim 3\times 10^{17}\,\rm eV $ (e.g. 
\cite{som01,hor03}), or even up to the ankle at 
$E_{\rm ank} \simeq 3\times 10^{18}\, \rm eV$, are produced in our
Galaxy (see e.g. \cite{nw00} for a recent review). 
 Meanwhile, CR acceleration to energies significantly exceeding 0.1~PeV with 
the conventional mechanism of nonrelativistic first-order shock acceleration 
by SNRs from typical (Type Ia and II)  
supernovae (SNe) is  problematic  \cite{lc83,bar99}. The origin of the knee
in the CR spectrum, in the form of a spectral-index break in the power-law 
all-particle spectrum by 
$\approx 0.3$ units at $E_1 \simeq 3\times 10^{15}\,\rm eV$, 
accompanied with a change in the CR composition, seems to suggest a new 
contribution to CRs in the Galaxy at these energies.

We have recently proposed a model \cite{wda04} that explains the entire CR 
spectrum  from GeV up to ultra-high energies (UHE) with a single  
population of sources, namely SNe.  It is important to realize 
that SNe consist of various types, not only the thermonuclear Type Ia 
SNe, but also the core-collapse Type II and Ib/c SNe. Observations
indicate that long-duration gamma-ray bursts (GRBs) are formed
by a subset
of Type Ib/c SNe that collapse to black holes. These black-hole formation 
events are observed as GRBs if the Earth happens to fall within the narrow 
opening angles of their relativistic beams (see \cite{mes02} for a  
review). 

GRBs have been proposed as effective accelerators of CRs
in the universe \cite{vie95,wax95,wax04}, and as probable sources of CRs up to 
ultra-high energies in our Galaxy \cite{der02,dh05}.   
Our model assumes that  CRs with energies below $\sim 100$ TeV are
produced in the conventional quasi-stationary scenario of continuous injection 
due to nonrelativistic shock acceleration by SNRs formed in 
all types of SNe, with subsequent modification of the source spectrum 
through energy-dependent 
propagation (see \cite{drury,gp85,ber90}). 
The principal proposal of our model is that up to the second 
knee, high-energy 
cosmic rays (HECRs) at $E\gtrsim 0.1\,$PeV are mostly
 due to a single (or a few) relatively recent Galactic GRB supernova 
event that  
occurred some $t_{0} \lesssim 10^{6}\,$yrs ago at distances 
$r_{0} \lesssim 1\,\rm kpc$ from us. UHECRs from extragalactic GRBs
dominate at $E \gtrsim E_2$.

We note that  a ``single-source" model has been proposed earlier by 
Erlykin and Wolfendale \cite{ew97}, who suggested that the knee could 
be due to a single ``normal" supernova event that occurred some 
$t\sim 10^4\,\rm yr$ ago within $r\sim 100\,\rm pc$ 
from us. Despite the apparent similarity in the approach, the differences 
between the GRB and SNR single-source models are substantial on both 
qualitative and quantitative levels. These include 
\begin{enumerate}
\item the possibility to
explain acceleration of particles up to ultra-high energies by the 
relativistic shocks formed by GRB outflows, which is very problematic in the 
case of SNRs formed in the collapse to neutron stars; and
\item the much larger total energy of HECRs injected, which permits the
source to have occurred at larger ($\sim 1\,$kpc) distances and from a 
significantly older GRB than for a single normal SN source. This makes it then 
easier to explain the likelihood of such an event, as well as the
low degree of anisotropy observed in HECRs. 
\end{enumerate}

Our model provides  a way to explain the origin and the sharpness 
of the knee at 
$E_1\simeq 3 \,\rm PeV$ as the consequence of pitch-angle 
scattering of CRs on the plasma waves injected in the interstellar medium 
(ISM) through dissipation of bulk kinetic energy of SNRs effectively 
on the pc-scale Sedov length. Furthermore, we also explain the origin of the
second knee in the spectrum of HECRs at $E_2 \simeq 4\times 10^{17}\,\rm eV$
 as an unexpected but reasonable consequence of  diffusive
propagation due to scattering with turbulence injected on 
a scale of 
$\sim 100\,\rm pc$. The latter corresponds to the  thickness of the 
Galactic disk, which therefore represents the maximum natural scale
for effective injection of plasma turbulence in the Galaxy. 
The transition from Galactic to extragalactic CRs occurs 
around and above the second knee. 

In this paper we present in more general terms the basic ideas and results 
of the model in Ref.\ \cite{wda04}, with particular emphasis on the effects of 
propagation and the allowed parameter space of the model.

\section{Propagation Effects and Spectrum of MHD turbulence}

\vspace{2mm}

Unlike relativistic electrons, relativistic protons and nuclei that 
contribute the bulk
of the measured CR energy density do not suffer significant 
radiative energy losses during their lifetime
in the Galaxy (heavier nuclei such as Fe can, however, experience substantial 
depletion through spallation). 
In particular, radiative synchrotron and Compton energy 
losses, which steepen the source spectra of relativistic electrons,
are entirely negligible for relativistic hadrons in the ISM and galactic halo, 
even at ultra-high energies.    
This admits only one remaining possibility to explain how the 
spectrum of cosmic rays with power-law index $\alpha \geq 2.7$ is steepened 
from the $\alpha_0 \approx 2.0$ -- $2.3$ injection indices
for CR source spectra predicted by first-order Fermi acceleration. 
Namely, one has to
invoke spectral steepening due to {\it energy-dependent} diffusive propagation 
in the interstellar medium \cite{gp85}. Spectral steepening
is then possible, but only if the energy density of CRs observed locally is 
higher than the energy density of CRs from 
``outside" the source injection region.  
The steep spectrum of the observed 
CRs, in particular HECRs,  implies that we are within a local ``bubble" of 
CRs, where the CR density is significantly higher than the mean energy 
density throughout the Universe.
At energies $E\gtrsim 5\times 10^{17}\,\rm eV$, the extragalactic 
cosmic rays have a larger energy density than the cosmic rays formed
within the Galaxy, so that the bulk of cosmic rays at higher energies
have a universal (extragalactic) origin. 
Spectral modifications due to 
 energy losses on cosmological time scales for UHECRs are then 
expected \cite{ber90}.

\begin{figure}
\centerline{
\epsfxsize=8.cm \epsfbox{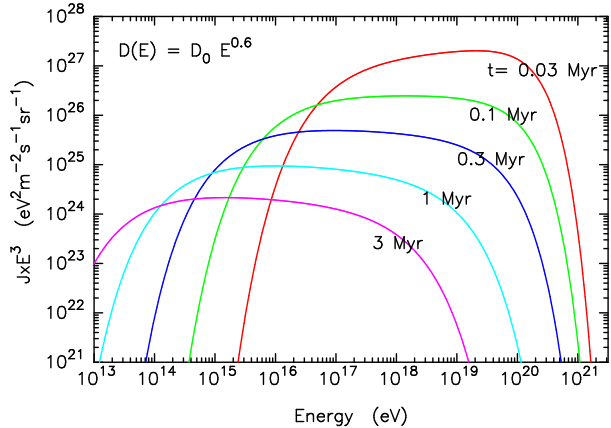}
}
\caption{ Time evolution of the energy distribution of CRs injected 
from a GRB at $r=1\,\rm kpc$ from an observer with power-law spectral index
$\alpha_0 = 2.2$, with a maximum (exponential cutoff) energy 
$E_{max}=10^{21}\,\rm eV$, and total energy $U = 10^{52}\,\rm ergs$. 
A diffusion coefficient $D(E)=D_0 E_{PeV}^{0.6}$ with 
$D_0 = 10^{29}\,\rm cm^2\, s^{-1}$ is assumed. 
}
\end{figure}

Our model assumes effective acceleration of CRs up to $\gtrsim 10^{20}
\,\rm eV$ by relativistic shocks of typical GRBs, one of which 
would, with reasonable probability, have 
occurred  in the Galaxy at a distance 
$r\lesssim 1\,\rm kpc$ from us within the last million years.
For spatially uniform diffusive propagation of cosmic rays
from a single impulsive burst-type source, the time evolution of the spectrum
$n(E;r,t)$ of CRs injected  
with initial energy distribution $N_0(E)$ at $t_0=0$ is given, in the absence of  energy 
losses, by the expression
\begin{equation}\label{single}
n(E,r,t)= {N_0(E) \over \pi^{3/2}\, r_{dif}^3}\;\exp[-(r/r_{dif})^2]\;
\end{equation} 
 \cite{syr59,aav95}.
Here  $D(E)$ is the  diffusion 
coefficient, and $r_{dif}$ is the energy-dependent diffusion radius of 
particles, given by 
\begin{equation}
r_{dif} \equiv r_{dif}(E,t)=
2 \sqrt{D(E)t}\;. 
\end{equation}
When an observer is inside the diffusion radius, that is, $r < r_{dif}(E,t)$, 
the steepening of the local hadron spectrum 
is the result of 
 larger volumes $V(E)\propto r_{dif}^3(E)$ occupied 
by particles of higher energies due to their faster diffusion. For 
a power-law diffusion coefficient $D(E)\propto E^\delta$, the source spectrum is steepened by 
a change in power-law index $\Delta\alpha \equiv \alpha- \alpha_0 = (3/2)\delta$. 
Note that the total spectrum of 
injected particles in the entire space does not change in this approximation. This
is easily checked by integration of Eq.\ (1) over $d^3r$. 
Without energy losses, the source spectrum is recovered 
if propagation  occurs in an effectively infinite volume.  
 
In Fig.\ 1 we show the time evolution of the spectrum of particles from 
a GRB at a distance $r=1\,\rm kpc$ from the observer. In this calculation,
the total energy 
of protons injected with power-law index $\alpha_0=2.2 $ is 
$U=10^{52}\,\rm ergs$. It is assumed that the maximum energy of accelerated 
particles is $E_{max}=10^{21}\,\rm eV$, and the diffusion coefficient is 
a single power-law with $\delta =0.6$ and $D_0=10^{29}\,\rm cm^2\, s^{-1}$.
The spectral steepening above energy $E$
at $t>t_{dif}(E)$, corresponding to $r_{dif}(E,t)>r$, results in 
$\alpha=\alpha_0 +(3/2)\delta =3.1$. A gradual decrease of the position 
of the exponential 
cutoff in the spectrum at the highest energies results from the assumption that 
particles leak from the Galactic halo on timescales $t_{esc}(E)$
that corresponds to their diffusion to $\approx 10$ kpc, so that 
$r_{dif}(E,t_{esc})=10\,\rm kpc$. This leads to an additional leaky-box 
type exponential modification of Eq.\ (1) by a factor $\exp[-t/t_{esc}(E)]$.

The magnitude of the CR flux at a given energy depends most importantly
on whether the low-energy exponential cutoff in Eq.\ (1) has reached this energy. Because
 $r_{dif}(E,t)\propto \sqrt{D(E)\,t}$ 
is the single parameter that defines the spectral evolution  in Eq.\ (1), the 
age of the source can be changed by assuming different absolute values of
the diffusion coefficient.

The CR diffusion in our model is due to pitch-angle scattering of 
protons and nuclei with magnetohydrodynamic (MHD) turbulence 
in the Galactic disk and halo. The spectrum of this turbulence, superposed on 
the Galactic magnetic field $B$, is described by spectral energy distribution 
$w(k)$ in wave number $k$. The shape of this spectrum is key to 
explain the origin of both knees in the HECR spectrum.

The Larmor radius of a CR ion with total energy 
$E=A m_p c^2 \gamma = E_{\rm PeV}\, \rm PeV $ 
and charge $Z$ propagating in a magnetic field of strength 
$B = B_{\mu{\rm G}}$~$\mu$G is 
\begin{equation}
r_{\rm L} (E) = {A m_pc^2 \beta\gamma \over Z eB}\sin\theta
\simeq {E_{\rm PeV} \over Z B_{\mu{\rm G}}}\;{\rm pc} \;,
\label{Larmor}
\end{equation}
where $\theta $ is the pitch angle, which we have supposed to be 
large enough ($\sim \pi/2$) so that $\sin\theta \sim 1$ on the right side 
in Eq.\ (\ref{Larmor}). Pitch-angle scattering on MHD waves takes place through 
a resonance between the ion gyration radius and the wavelength $k^{-1}$,
i.e. $k \,r_{\rm L} \sim 1$ (see,
e.g., \cite{dml96,sch02} for more detailed treatments). This results in 
the mean-free-path $\lambda(E)$, and hence the diffusion coefficient 
$D(E)=c\lambda/3$ of CRs with energy $E$, being tightly connected with
the energy density in the MHD spectrum at $k\sim r_L^{-1}$, so that 
$\lambda = r_{\rm L}U_B /\bar k w(\bar k)$ \cite{drury}. Therefore
a local power-law index $q$ in the turbulence spectrum 
$w(k)\propto k^{-q}$ near wavenumber $k_0=1/r_{\rm L}(E_0)$ translates
to a power-law diffusion coefficient $D(E)\propto E^{2-q}$ in the vicinity of $E_0$. 

\begin{figure}
\centerline{
\epsfysize=5.cm \epsfbox{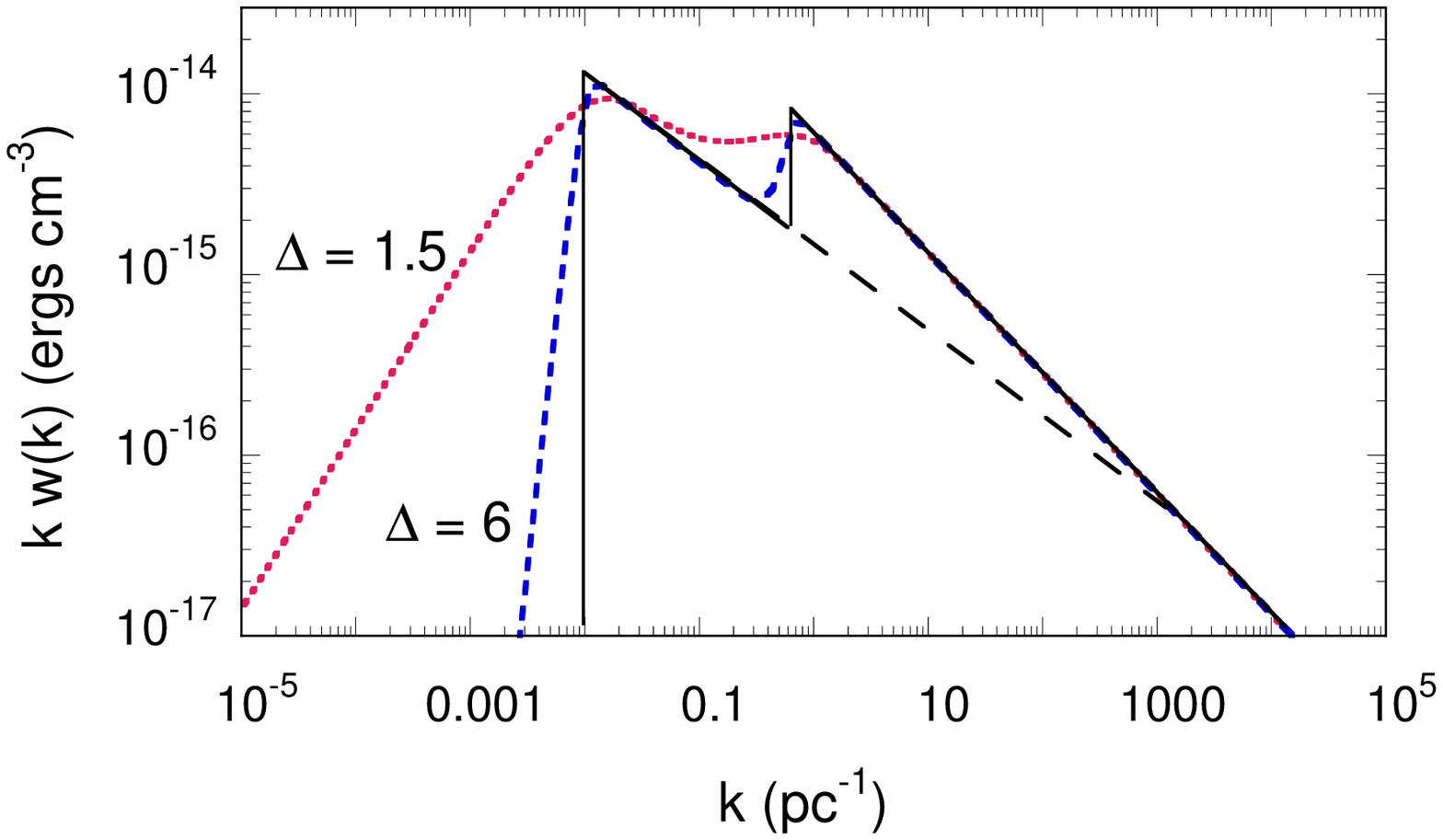}
\hspace{7mm} 
\epsfysize=5.3cm \epsfbox{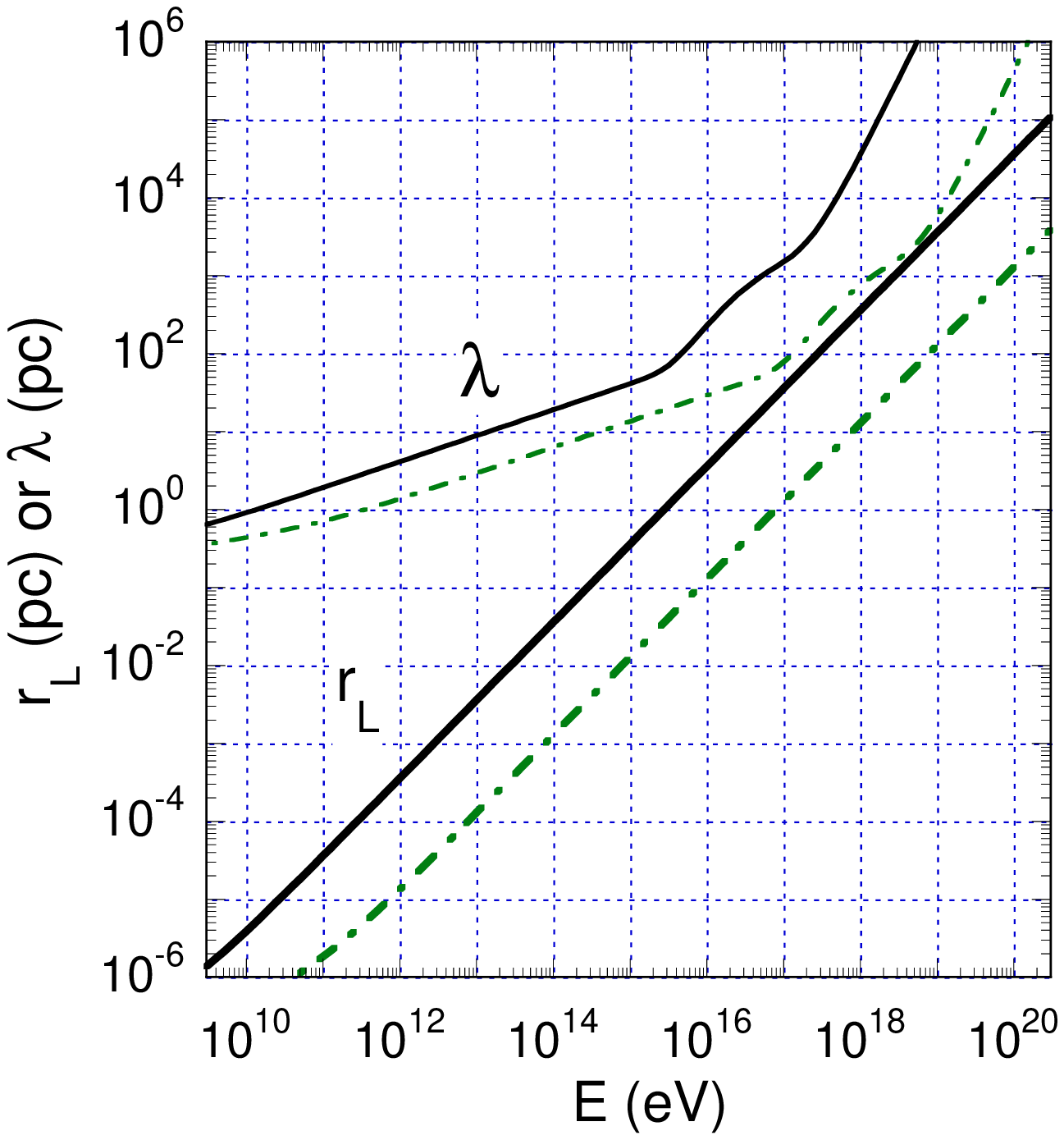}
}
\caption{ {\protect {\bf (a)} {\it (left)}} 
Wave turbulence spectrum used to model 
CR propagation in the Galaxy, assuming injection of turbulence at
scales $k_1=1/1.6\,\rm pc^{-1}$ and $k_2= 0.01\,\rm pc^{-1}$, followed
by cascading of turbulence to smaller size scales and larger wave numbers. 
An idealized model is shown by the solid lines and, after
smoothing, by the dotted and short-dashed curves for the smoothing
parameter $\Delta = 1.5$ and 6, respectively (see \cite{wda04} for details).
~~{\protect {\bf (b)} {\it (right)}} Larmor radius 
$r_{\rm L}(E)$ and the mean-free-path $\lambda$ of CR protons (solid curves)
 and Fe nuclei ($Z = 26, A = 56$; dot-dashed curves) with total energy
 $E = A\gamma m_pc^2$ in a magnetic field with mean strength of $3 ~\mu$G.
}
\end{figure}

Fig.\ 2a shows the spectrum of MHD turbulence used in Ref.\ \cite{wda04}
for calculations of $\lambda(E)$.  
Here we assumed that MHD turbulence is injected into the Galactic disk and halo 
by two different processes on two distinct size scales.   Injection on the 
$k_{1}^{-1} \sim 1$ pc scale is likely 
due to SNRs after reaching 
the Sedov phase.  Turbulence injected at scales 
$k_{2}^{-1}\sim 100$ pc, which is of the order of the characteristic 
thickness of the gas disk of our Galaxy, may be due 
to the interaction of high velocity clouds with the Galactic disk. 
The solid and long-dashed lines 
correspond to the power-law indices $q=5/3$ and $q=3/2$ for the Kolmogorov and 
Kraichnan turbulence spectra, respectively, after 
further cascading of the injected MHD waves from these
2 types of sources. 
The total spectrum of turbulence that results after application 
of a smoothing procedure with parameters $\Delta=1.5$ and $\Delta=6$ 
(see \cite{wda04} for details) are shown in Fig.\ 1a by dotted and short 
dashed curves, respectively.

The mean free path for scattering of CR protons and ion nuclei,
calculated for the plasma turbulence spectrum plotted in Fig.\ 1a 
(for $\Delta=1.5$), are
shown in Fig.\ 2b. As is apparent from Eq.\ (\ref{Larmor}), for the characteristic Galactic 
magnetic field $B\simeq 3\,\rm \mu G$, the MHD turbulence near
the break in the $w(k)$ spectrum
at $k_1\sim 1\,\rm pc^{-1}$ resonates with protons with 
energies $E_1\sim 3\,\rm PeV$. This explains the origin of the first knee. The second break 
in the spectrum of turbulence at $k_2\sim 1/(100 \,\rm pc)$ is the cause of 
the second knee at $E_2 \sim 3\times 10^{17}\,\rm eV$ in the CR spectrum.  
For particles of higher energies, there is not sufficient energy in resonant
MHD waves 
to prevent rapid escape from the Galactic halo. 
This is seen in Fig.\ 2b as a very steep increase of
$\lambda$ with energy $E$ above $\sim 2\times 10^{17}\,\rm eV$ for the
protons. The power-law spectrum of the MHD waves at 
$k<k_2$ in Fig.\ 2a (dotted curve) could be as hard as $q\simeq 0$. 

One of the major objections to a model where the transition between
the galactic and extragalactic components occurs in the vicinity of 
the second knee is that fine-tuning is required to smooth
the transition where the galactic component
exponentially cuts off and the extragalactic component emerges. 
This criticism is ameliorated in this model because no exponential 
cutoff of the galactic component is required for a propagation model, 
as compared to model where the maximum energy of the galactic CR 
source is due to acceleration and loss or escape processes.
Depending on the wave turbulence spectrum at $k \lesssim 0.01$ pc$^{-1}$,
a break to a softer power-law may occur due to propagation effects at
the second knee. This will naturally smooth any transition to a second
component.

\section{CRs from Local GRBs and Propagation Parameters} 

\vspace{3mm}

Our model predicts a transition from the Galactic to extragalactic 
components near and above the second knee in the all-particle spectrum
at $E\sim (3$ -- $5)\times 10^{17}\,\rm eV$.  
For CR nuclei with larger $Z$, the positions of both knees move to 
higher energies because of the smaller gyroradii for the same total particle energy, 
as implied by Eq.\ (\ref{Larmor}).  Injection of HECRs from a single GRB source 
also allows, as demonstrated in \cite{wda04},
a good fit to the
cosmic-ray ion spectra measured with KASCADE (Karlsr\"uhe Air
Shower Array) \cite{hor03,kam01} through
the first knee of the cosmic ray spectrum  at energies $\approx 10^{14}$ eV -- $10^{17}$ eV.

\begin{figure}[b]
\centerline{
\epsfxsize=8.cm \epsfbox{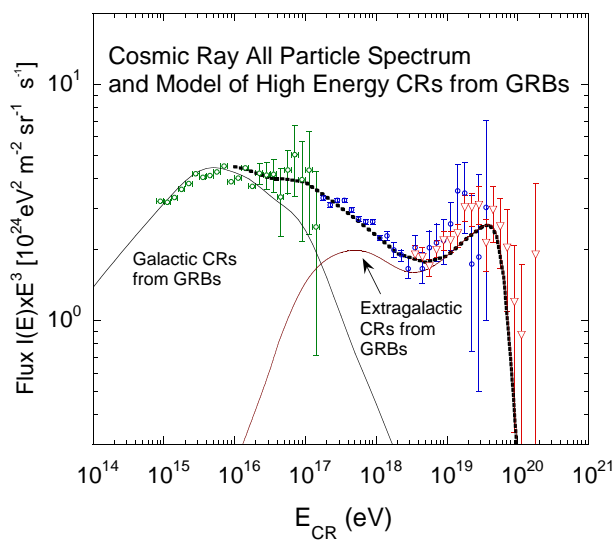}
}
\caption{ The total spectrum of HECRs above 100 TeV contributed by a 
single recent GRB (timescale $t\sim 1\,$Myr), and by extragalactic sources 
at energies $\gtrsim 10^{17}\,\rm eV $ on cosmological timescales 
(see Ref.\ \cite{wda04}).}
\end{figure}

 The total all-particle spectrum, including 
both galactic and extragalactic CR components, is shown by the dotted curve
in Fig.\ 3. The steepening of the all-particle spectrum from
a power law with $\alpha \approx 2.7$ to one with
$\alpha \simeq 3$ would imply 
steepening in the  index $\delta$ of
the diffusion coefficient by $\Delta\delta = (2/3) \Delta \alpha \simeq 
0.2$. Assumption of a Kolmogorov spectrum for $w(k)$ at 
$k > k_1\sim 1\,\rm pc^{-1}$ results in $\delta_1 =2-5/3=1/3$, which whould
imply $\delta_2 \approx 0.53 $ at energies above the first knee. The
latter value is
very close to the index $\delta_2=2-3/2=0.5$ resulting from the assumption
of Kraichnan turbulence at $k_2\leq  k\leq k_1$ \cite{wda04}. However, the 
true Kraichnan-type evolution of the turbulence would then overtake the 
Kolmogorov turbulence at $k\gg k_1$ as well.

The spectrum of turbulence formed at scales smaller 
than those of the active injection scale
may not be the result of turbulent cascades to smaller scales for 
the turbulence injected at $\sim 100\,\rm pc$ scales, but may rather  
 reflect the 
{\it rates of injection} of turbulent MHD energy at different spatial scales between
$\sim 1\,\rm pc$ and $\sim 100\,\rm pc$. For example, considering even 
only SNe as sources of MHD turbulence, we note that the sizes of 
SNR shells reach $\gtrsim 10\,\rm pc$, such as in W44 or W50. 
Thus, although in Figure 2a
we have assumed that SNRs inject MHD turbulence only at scales 
$\sim 1\,\rm pc$, in reality 
their kinetic energy is dissipated in the interstellar medium 
(ISM) through much larger scales. Similar wide range of
spatial scales should be also expected from other potential sources
of MHD turbulence, such as high velocity clouds or large scale bubbles blown 
by stellar winds and multiple SNe in regions of active star formation.  

The spectrum of turbulence at $k>k_1$ could also be of 
the Kraichnan form
$q\simeq 3/2$, resulting in $D(E)$ with $\delta \simeq 0.5$. This
could be more preferable to explain the lower-energy part of
the CR spectrum for continuous injection of 
CRs from SNRs. In this case, the observed spectrum has an index  
$\alpha = \alpha_0 +\delta$, which implies a reasonably hard source
spectrum with $\alpha_0 \simeq 2.2$. 
The break at the knee by $\Delta \alpha \simeq 0.3$ in this case would
imply a characteristic spectral index $q\simeq 1.3$ for 
$w(k)\propto k^{-q}$ at $k_2 < k \leq k_1$. It would also
suggest a more 
uniform injection of turbulence over all length scales from $\simeq 1\,$pc
to $\sim 100$\,pc in the ISM. In this  
scenario, it is important to realize that
 the diffusion coefficient at energies between  
the two knees would correspond to $\delta \simeq 0.7$, resulting in the
steepening of the single-source spectrum by $\Delta \alpha \simeq 1$. 
Thus, this scenario suggests a rather hard injection spectrum  of
CRs by GRBs, namely $\alpha_0 \simeq 2 $-$2.2$. The latter value is allowed
if we take into account that the real spectrum of HECRs from the 
local GRB above the first knee could  easily be in the range of
$\alpha \sim 3.2$ if one includes a small
 contribution of extragalactic HECRS at 
$E\gtrsim 10^{16}\,\rm eV$ (see Fig.\ 3).

Thus, both Kolmogorov and Kraichnan types of turbulence at scales 
$k> k_1$ are possible. In the latter case an additional small 
steepening of the single-source spectrum below the first knee 
due to the relative proximity of the low-energy exponential 
turnover (see Fig.\ 1)  can also be invoked. In any case, the 
sharpness of the spectral 
break at the knee directly reflects the sharpness of the 
spectral break in $w(k)$ at $k_1$. 

The important model parameters are the age of the GRB and its distance. 
In the spectral fit shown in Fig.\ 3, we have used $t=0.2\,\rm Myr$ and
$r=0.5 \,\rm kpc$, with $D_0 \equiv D(1\,\rm PeV) = 1.5\times 10^{30}$ cm$^2$ s$^{-1}$.
The latter is calculated assuming the ratio
of MHD to Galactic magnetic field energy densities $\xi = 0.03$.
For these parameters, protons with $E=1\,\rm PeV$ diffuse to scales 
$r_{dif}(1\,\rm PeV) \simeq 2\,\rm kpc$, i.e., well beyond the 
assumed distance to the source. Thus, we could also assume a distance
to the source $r \sim 1\,\rm kpc$ for the same total energy 
($10^{52} \,$ergs), while keeping the exponential turnover of the spectrum 
at low energies significantly below 1\,PeV.  
Because of the invariance of the local 
CR spectrum with respect to the product $D_0 \times t$, we could equally 
assume that the 
diffusion coefficient was smaller by one order 
of magnitude (with $\xi \sim 0.3$), but a GRB age
$\gtrsim 1\,\rm Myr$. The smaller diffusion coefficient is also  
preferable because in that case the diffusion timescale  
of $\sim 10\,$GeV CRs out of the galactic disc would be in better 
agreement with the generally accepted value $t_{esc} \sim 10^7\,\rm yr$ 
for these energies.

\begin{figure}
\centerline{
\epsfysize=7cm \epsfbox{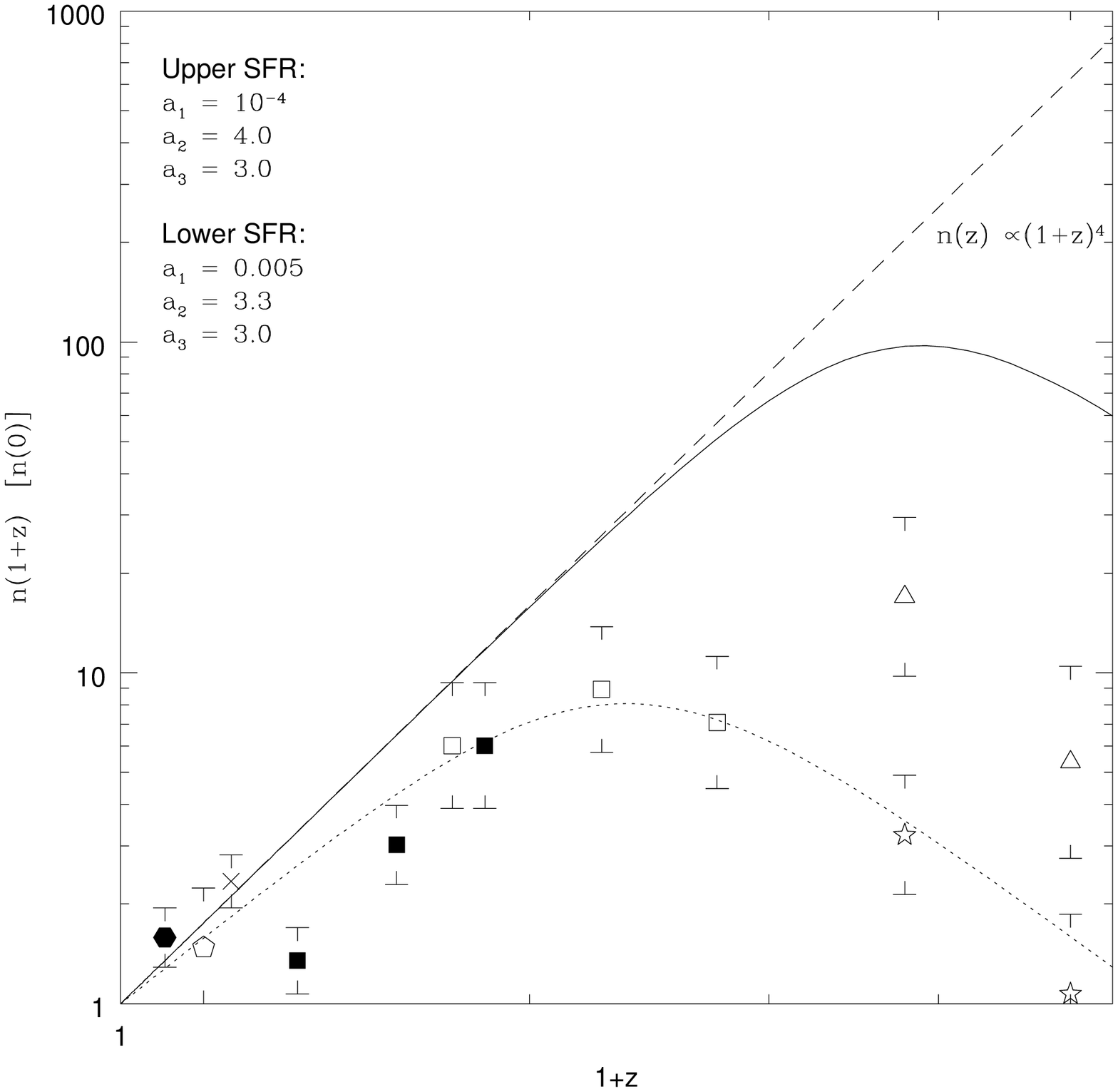} \hspace{1cm}
\epsfysize=6.9cm \epsfbox{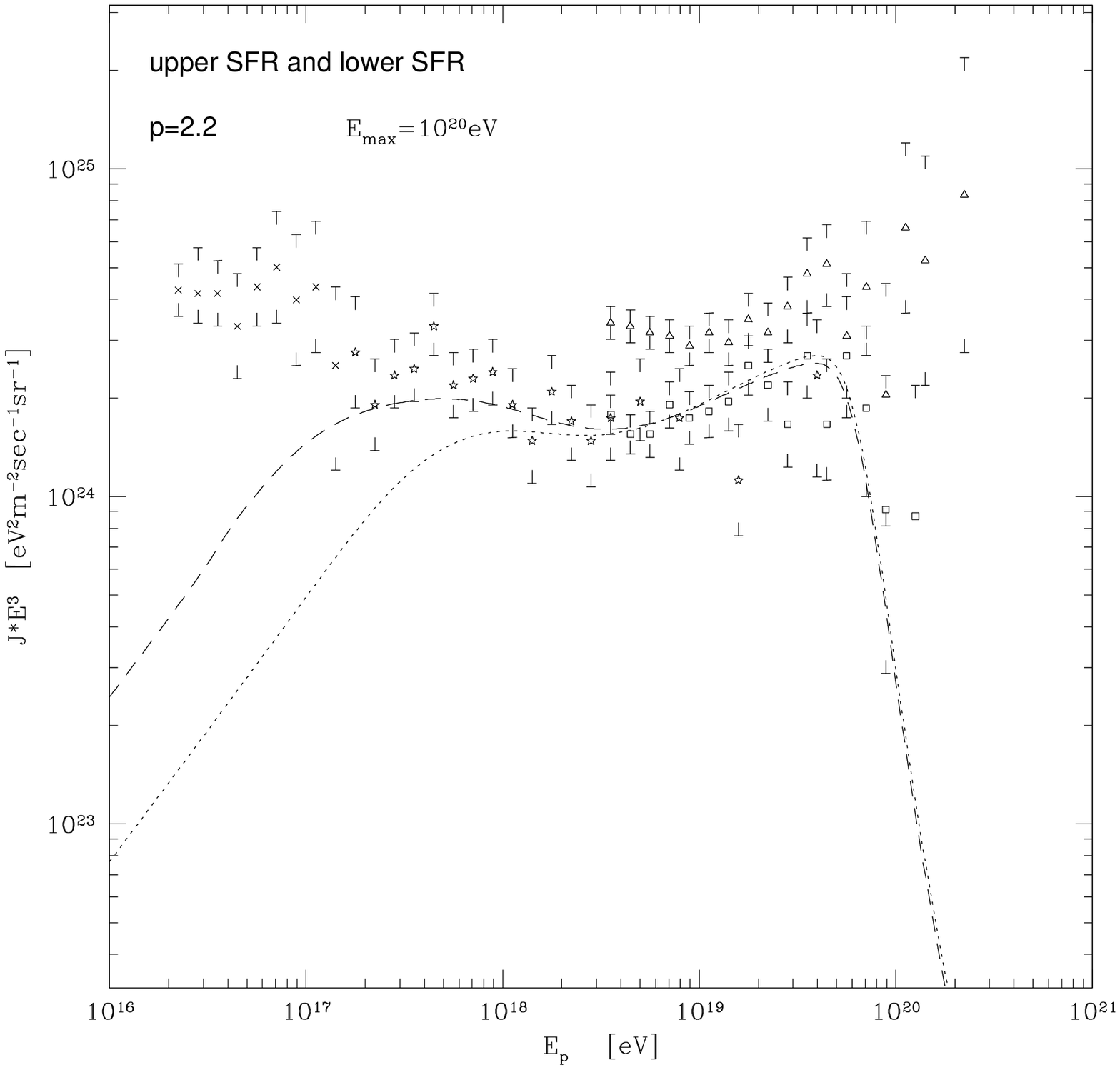}
}
\caption{{\protect \bf (a)} ({\it left}) 
The history of evolution of the star formation rate (SFR) in the universe 
as a function of redshift $1+z$,
normalized to the current SRR.  
 The dotted curve shows the lower limit to the SFR evolution implied
by measurements of the blue and UV energy density, and the solid 
curve shows the SFR
corrected for dust extinction (see \cite{wda04} for detailed discussion). 
The dashed line displays the relation
$n(z)=n(0)(1+z)^4$ used by \cite{ber05} for calculations of
the fluxes of extragalactic CRs. 
{\protect \bf (b)} ({\it right}) Calculated fluxes of extragalactic CRs 
assuming that the injection of UHECRs in the universe was due to 
GRBs with a rate density proportional to the 
 minimum (dotted curve) and maximum (solid curve) SFR functions 
shown in Figure 4a. 
Note that the spectra are not normalized to each other at high
energies. Instead, the normalization for both of them corresponds to
the same value for the current ($z=0$) injection rate.    
}
\end{figure}

A larger distance and age can help to
improve the likelihood of a local GRB. As estimated in \cite{wda04},
the rate of occurrence of a GRB in our Galaxy is 
$\approx 1$ per $10^4\,$yr, in agreement also with the estimates in
\cite{vmg03}. Using this, the mean number of GRBs that
would occur in the Galactic disc at $r=r_{\rm kpc}\,\rm kpc$ from us    
during $t=t_{\rm Myr} 10^6\,$yr is estimated as 
$\overline{N}_{\rm GRB} \simeq (0.45$-$1.3)\, r_{\rm kpc}^{2} \, 
t_{\rm Myr}$.  

Another advantage of a longer age of the local GRB as large as $\sim 1\,\rm Myr$
is the possibility to explain better the observed small anisotropy 
$\omega =(J_{max}-J_{min})/(J_{max}+J_{min})\sim (0.15\pm 0.05)\%$ 
\cite{wat84,hil84} in the knee region (the anisotropy is, however,
increased for a more distant source).
For the spatial distribution of the CRs $n\equiv n(E,r,t)$ given by 
Eq.\ \ref{single}, calculations of the anisotropy \cite{gp85} result in 
\begin{equation}\label{anisotropy}
\omega = \frac{3D}{cn} \frac{\partial n}{\partial r} =  
\frac{3 r}{ 2 c t} \cong \frac{0.4 \, r_{\,\rm kpc}}{ t_{Myr}}\, \%.
\end{equation}
For $r = 0.5$ kpc and $t=2\,$Myr the anisotropy can be as small as 0.1\%.
An interesting point in Eq.\ (\ref{anisotropy}) is that $\omega$ 
is independent of energy for a spherically symmetric single-source model.

\section{Extragalactic Cosmic Rays}

\vspace{3mm}

The rapid decline of the CR flux from local GRBs above the second knee results in
the  contribution of extragalactic component in the all-particle spectrum
dominating near and above the second knee. Calculations of the extragalactic component 
as shown in Fig.\ 3 for our model of CRs from GRBs includes photomeson interactions, 
$e^+-e^-$ pair production, and adiabatic cooling of UHECRs \cite{ber90}. 
We assume that the rate density of GRBs
is proportional to the cosmological 
 star formation rate (SFR) history of the universe \cite{madau}. 
For the two rates shown in Fig.\ 4a that correspond to minimum and 
maximum SFRs, calculations in \cite{wda04} result in the two spectra for the 
extragalactic component shown in Fig.\ 4b. An interesting result here is
that in the framework of this model, the ankle in the spectrum of CRs observed
at $E\simeq 3\times 10^{18}\,\rm eV$ is formed in the process of cooling of
UHE protons on cosmological timescales. 

Similar spectral behavior for the extragalactic CR component at 
$E\geq 10^{18}\,\rm eV$ as shown  in Fig.\ 4b,
where the ankle is explained
as a consequence of photopair losses of UHECRs formed 
at high redshift, 
were also derived by Berezinsky and collaborators
(e.g. see \cite{ber04,ber05} and references therein). 
These authors \cite{ber05} consider a model
where UHECRs are accelerated by active galactic nuclei
and assume
 cosmological evolution of the injection rate of UHECRs  
$\propto (1+z)^4$ (see Fig.\ 4a).
It remains to be studied if these two principal 
options (GRBs and AGNs) for the sources of UHECRs in the universe can be distinguished
from each other observationally as a result of differences in their
evolutionary histories. 

\begin{figure}
\centerline{
\epsfxsize=8.5cm \epsfbox{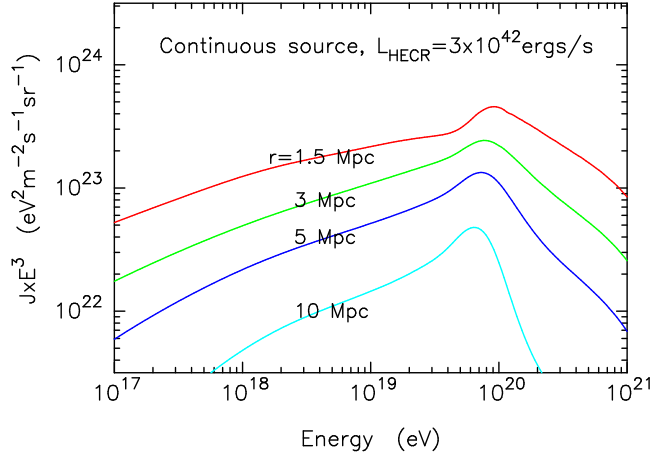}
}
\caption{Fluxes of UHECRs from a local, continuous extragalactic source 
population at
distance $r$ that injects HECRs into the intergalactic medium with  
spectral index  $\alpha_0 = 2.2$, a maximum ('exponential cutoff') energy 
$E_{max}=10^{21}\,\rm eV$, and total power 
$L_{HECR}(E\geq 1\,\rm PeV) = 3\times 10^{42} \,\rm ergs\, s^{-1}$.  
Diffusive propagation with $\delta =0.5$ and 
$D(10^{19}\,\rm eV)=10^{34}\,\rm cm^2\, s^{-1}$ in the local intergalactic 
medium is assumed.
}
\end{figure}

The spectra of UHECRs resulting from injection of UHECR protons in the universe
on cosmological timescales show a sharp (``GZK") cutoff above the GZK energy
$E\simeq 6\times 10^{19}\,\rm eV$. The UHECR spectrum in Fig.\ 3 
(or Fig.\  4b) agrees with the 
HiRes data, but is in disagreement with the AGASA results at 
$E\sim 10^{20}\,\rm eV$. If Auger observations  show any significant 
excess over the exponential GZK cutoff at these energies, this would
imply that there are other recent ($\lesssim 10^{8}\,\rm yr$) local
source sources of extragalactic origin in our vicinity at
$\lesssim 10\,\rm Mpc$ that produce this flux. One 
possibility is that the excess would be due to cosmic ray
ions (e.g., \cite{all05}).

In the framework of our model,
 such extragalactic sources could be connected
with starburst galaxies in the local group, such as M82 and NGC 253,
both at distances $r \sim 3.5\,\rm Mpc$. Taking into account that
the supernova rate in these galaxies is about 0.3 -- 1 per year, and that
the estimated GRB rate in our Galaxy is about (0.3 -- 1)\% of the 
supernova rate, the mean GRB rate in the starburst galaxies is 
estimated as one per $\sim 300$ -- 1000 yrs. If  
the total energy of CRs accelerated by a typical GRB is indeed about 
$10^{52}\,\rm ergs$, as for our local Galactic GRB, the characteristic
injection power of UHECRs from starburst galaxies averaged over the
timescale of $\sim 10^8$\,yr can be $\sim (1-3) \times 10^{42}\,\rm 
ergs\,s^{-1}$. In Fig.\ 5 we show the UHECR fluxes expected from a single
{\it continuous} source (which is valid for a GRB model because 
of the large number of GRBs from these starburst galaxies
within the last 100 Myrs) at a distance $r$ from us. The fluxes are calculated
in the framework of a diffusion propagation model from a single source,
assuming a diffusion coefficient with $\delta = 0.5$ normalized at 
$D(10^{19}\,\rm eV)=10^{34}\,\rm cm^{2} s^{-1}$. Note that the Larmor
radius of a $10^{19}\,\rm eV$ proton in the magnetic field 
$B_{extragalactic} \sim 10^{-7}\,\rm G$ would be about
$3\times 10^{23}\,\rm cm$. This implies that the assumed diffusion 
coefficient would still be larger, by a factor of 3, than for Bohm
diffusion. The assumption of a different propagation model (or diffusion 
coefficient) in the intergalactic space would change the fluxes shown in 
Fig. 5, and will require a separate study.

\section{Conclusions}

\vspace{3mm}

We have described a complete model for cosmic rays comprising 
a single type of sources, namely SNe. Because of the 
wide diversity of SNe types, the efficiency of 
CR acceleration varies dramatically from  
Type Ia and II SNe, with SN ejecta speeds in the range
of 3,000 -- 30,000 km s$^{-1}$, to Type Ib/c SNe, with 
SN ejecta reaching highly relativistic velocities in the subset
of Type Ib/c SNe that collapse to black holes and form GRBs.
Because GRBs are found in our Galaxy, we expect that CRs 
to the highest energy will also be accelerated by past GRB sources
in the vicinity of Earth. 

The transition from the Galactic to the extragalactic component
occurs at the second knee and, as also suggested by Berezinsky and collaborators \cite{ber04}, 
the ankle is a consequence of pair-production interactions
(similar conclusions have been reached in Ref.\ \cite{lemoine}, 
but without proposing a specfic source model for the high-energy CRs). As we 
show \cite{wda04,da03}, the large energy in CRs in a GRB makes
it likely that GRBs will be detectable high-energy neutrino sources
with IceCube \cite{hal}. Detection of even one PeV neutrino coincident with
a GRB will confirm that GRBs are efficient accelerators of 
high-energy cosmic rays and will support this model for cosmic-ray 
origin.

\vspace{3mm}

\section*{Acknowledgements}
We would like to acknowledge here the contribution of Dr. Stuart Wick
to development of this model in our earlier work \cite{wda04}.
The work of C.\ D.\ is supported by the Office of Naval
Research and the NASA {\it Gamma Ray Large Area Space Telescope}
(GLAST) program.


\vspace{3mm}

\section*{References}

\begin{thebibliography}{}


\bibitem{drury} Drury, L. O'C., 1983, Rep. Prog. Phys. {\bf 46}, 973 


\bibitem{gp85} 
Ginzburg, V.\ L.\, Ptuskin, V.\ S.\, 1985, Astrop. Sp. Phys. Rev.\ {\bf 4}, 
161


\bibitem{som01} Sommers, P., in: 27th International Cosmic Ray Conference 
(Copernicus Gesellschaft, Hamburg, Germany) Invited, Rapporteur, and 
Highlight Papers (2001) 170.

\bibitem{hor03}
H\"orandel, J. R., 2003, Astropart. Phys. {\bf  19}, 193

\bibitem{nw00} Nagano, M., Watson,  A.\ A., 2000, Rev.\ Mod.\ Phys.\
{\bf 72}, 689

\bibitem{lc83}
Lagage, P. O., Cesarsky, C. J., 1983, Astron. Astrophys. {\bf 125}, 249.

\bibitem{bar99}
Baring, M. G., Ellison, D. C., Reynolds, S. P., Grenier, I. A., Goret, P.,
1999, Astrophys. J. {\bf 513}, 311 

\bibitem{wda04} Wick, S.\ D., Dermer, C.\ D., and Atoyan, A., 2004, Astropart.\ Phys., 21, 125

\bibitem{mes02} M\'esz\'aros, P., 2002, Ann.\ Rev.\ Astron.\ Astrophys.\ 
{\bf 40}, 137.

\bibitem{vie95} Vietri, M., 1995, Astrophys.\ J.\ {\bf 453}, 883.

\bibitem{wax95} Waxman, E. 1995, Phys.\ Rev.\ Letters {\bf 75}, 386.
\bibitem{wax04} Waxman, E.\ 2004, Astrophys.\ J.\ , {\bf 606}, 988. 

\bibitem{der02} Dermer, C.\ D.\ 2002, Astrophys.\ J., 574, 65.
\bibitem{dh05} Dermer, C.~D., and Holmes, J.~M.\ 2005, Astrophys.\ J.\ Lett., 628, L21 

\bibitem{ber90} 
Berezinsky, V. S., Bulanov, S. V., Dogiel, V. A., Ginzburg, V. L.,  Ptuskin, 
V. S., 1990, Astrophysics of Cosmic Rays (Amsterdam: North Holland) 


\bibitem{ew97}
Erlykin, A. D., and Wolfendale, A. W. 1997, J. Phys. G: Nucl. Part. Phys.
 {\bf 23}, 979 

\bibitem{syr59}
Syrovatskii, S. I., 1959, Sov. Astron. AJ {\bf 3}, 22
\bibitem{aav95} Atoyan, A.\ M., Aharonian,  F.\ A., and V\"olk, H.\ J.,  1995,
Phys.\ Rev.\ {\bf D 52}, 3265.
%
\bibitem{dml96} Dermer, C.\ D., Miller, J.\ A.\, Li, H., 1996, Astrophys.\ J.\
{\bf 456}, 106. 
%
\bibitem{sch02}  Schlickeiser, R., 2002, Cosmic Ray Astrophysics (Springer-Verlag, 
Berlin), chpts.\ 14 and 17.
\bibitem{kam01} Kampert K.-H.\ et al., 2001, in: 27th International Cosmic Ray Conference (Copernicus Gesellschaft, Hamburg, Germany) Invited, Rapporteur, and Highlight Papers, p.240
\bibitem{vmg03} Vietri, M., De Marco, D., Guetta, D. 2003, Astrophys.\ J.\ 
{\bf 592}, 378 
\bibitem{wat84} Watson, A. A., 1984, Adv.\ Space Res.\ {\bf 4}, 35.
\bibitem{hil84} Hillas, A. A., 1984, Ann.\ Rev.\ Astron.\ Astrophys.\ 
{\bf 22}, 425
\bibitem{madau} Madau, P., Pozzetti, L., Dickinson, M., 1998, 
Astrophys.~J.\ {\bf 498}, 106.

\bibitem{ber04} Berezinsky, V., 
Gazizov, A., \& Grigorieva, S.\ 2004, Nuclear Physics {\bf B}, Proc.\ 
Supp., {\bf 136}, 147. 

\bibitem{ber05} Berezinsky, V.\ S., Gazizov, A.\ Z., Grigorieva, S. I., 
2005, Phys. Lett. {\bf B 612}, 147, see also astro-ph/0210095.
 
\bibitem{all05} Allard, D., et al.\ 2005, submitted to Astronomy and Astrophysics 
(astro-ph/0505566).

\bibitem{lemoine} Lemoine, M.\ 2005, Phys.\ Rev.\ {\bf D 71}, 
083007.

\bibitem{da03} Dermer, C.\ D., and  Atoyan, A.\ 2003, Phys.\ Rev.\ Letters {\bf 91}, 071102.

\bibitem{hal} Halzen, F.\ 2005, Nuclear 
Physics {\bf B} Proc.\ Supp., {\bf 145}, 301 

 \end{thebibliography}
\end{document}